\documentclass[%
 aip,
 amsmath,amssymb,
 reprint,%
]{revtex4-1}

\usepackage{graphicx}% Include figure files
\usepackage{dcolumn}% Align table columns on decimal point
\usepackage{bm}% bold math
\usepackage[utf8]{inputenc}
\usepackage[T1]{fontenc}
\usepackage{mathptmx}
\usepackage{etoolbox}
\usepackage[normalem]{ulem}

\makeatletter
\def\@email#1#2{%
 \endgroup
 \patchcmd{\titleblock@produce}
  {\frontmatter@RRAPformat}
  {\frontmatter@RRAPformat{\produce@RRAP{*#1\href{mailto:#2}{#2}}}\frontmatter@RRAPformat}
  {}{}
}%
\makeatother
\usepackage{xcolor}
\begin{document}

%jack\preprint{AIP/123-QED}

\title[]{The rise and fall of stretched bond errors: Extending the analysis of Perdew-Zunger self-interaction corrections of reaction barrier heights beyond the LSDA}
% Force line breaks with \\
\author{Yashpal Singh}
\affiliation{Department of Physics, Central Michigan University}%Lines break automatically or can be forced with \\
\author{Juan E. Peralta}%
\affiliation{Department of Physics and Science of Advanced Materials PhD Program, Central Michigan University}
\author{Koblar A. Jackson}
\email{jacks1ka@cmich.edu}
\affiliation{Department of Physics and Science of 
Advanced Materials PhD Program, Central Michigan University}

\date{\today}% It is always \today, today,

\begin{abstract}
Incorporating self-interaction corrections (SIC) significantly improves chemical reaction barrier height predictions made using density functional theory methods. We present a detailed, orbital-by-orbital analysis of these corrections for three semi-local density functional approximations (DFAs) situated on the three lowest rungs of the Jacob's Ladder of approximations.  The analysis is based on Fermi-L\"owdin Orbital Self-Interaction Correction calculations performed at several steps along the reaction pathway from the reactants (R) to the transition state (TS) to the products (P) for four representative reactions selected from the BH76 benchmark set. For all three functionals, the major contribution to self-interaction corrections of the barrier heights can be traced to stretched bond orbitals that develop near the TS configuration.  The magnitude of the ratio of the self-exchange-correlation energy to the self-Hartree energy  (XC/H) for a given orbital is introduced as an indicator of one-electron self-interaction error. For the exact, but unknown density functional, XC/H = 1.0 for all orbitals, while for the practical DFAs studied here, XC/H spans a range of values.  The largest values  are obtained for stretched or strongly lobed orbitals.  We show that significant differences in XC/H for corresponding orbitals in the R, TS, and P configurations can be used to identify the major contributors to the SIC of barrier heights and reaction energies. Based on such comparisons, we suggest that barrier height predictions made using the SCAN meta-generalized gradient approximation may have attained the best accuracy possible for a semi-local functional using the Perdew-Zunger SIC approach.
\end{abstract}

\maketitle

% \begin{quotation}
% The ``lead paragraph'' is encapsulated with the \LaTeX\ 
% \verb+quotation+ environment and is formatted as a single paragraph before the first section heading. 
% (The \verb+quotation+ environment reverts to its usual meaning after the first sectioning command.) 
% Note that numbered references are allowed in the lead paragraph.
% %
% The lead paragraph will only be found in an article being prepared for the journal \textit{Chaos}.
% \end{quotation}

\section{INTRODUCTION}
Density functional theory (DFT) is widely used for computing the properties of molecules and solids because it offers useful accuracy at much greater efficiency than competing wave function methods.  The recently developed strongly constrained and appropriately normed (SCAN) functional\cite{sun2015strongly} and the related r$^2$SCAN functional\cite{furness2020accurate} yield remarkable accuracy for many properties\cite{sun2016accurate,shahi2018accurate,zhang2018efficient} by satisfying all of the 17 constraints known for a semi-local density functional approximation (DFA).  One constraint that cannot be satisfied by a semi-local functional is one-electron self-interaction freedom, the fact that, in a one-electron system, the electron interacts only with an external potential and not with itself.  The self-interaction error (SIE) that arises as a result of a DFA not satisfying this constraint remains a pervasive and elusive challenge in DFT.  
There has been significant recent interest in the impact of electron self-interaction on the performance of DFAs in calculating chemical reaction barrier heights,\cite{shukla2023,zhang1998challenge, Lynch01, Zhao2005, Xu2011, Branko97, Shahi19}  where semi-local DFAs are known to underestimate barriers.\cite{patchkovskii2002improving,grafenstein2004impact,porezag2005,janesko2008hartree}  Mishra \textit{et al.}\cite{mishra2022study} used the Fermi-L\"owdin orbital self-interaction correction (FLOSIC) method\cite{pederson2014communication,pederson2015fermi,yang2017full} with the local density approximation (LDA) of Perdew and Wang\cite{perdew1992accurate} for both the Perdew-Zunger self-interaction correction\cite{perdew1981self} energy functional (PZSIC) and the locally scaled self-interaction correction of Zope \textit{et al.}\cite{zope2019step} (FLO-LDA and LSIC-LDA, respectively) to compute barrier heights for the BH76\cite{zhao2005multi,zheng2009dbh24} benchmark set of reaction barriers.  They showed that both FLO-LDA and LSIC-LDA give dramatically improved predictions of barrier heights relative to LDA.  They also estimated density-driven and functional-driven contributions to the error, concluding that the error made by LDA is mostly functional-driven.  Kaplan and co-workers\cite{Kaplan2023} also used the FLOSIC method and PZSIC to compute BH76 barriers for the Perdew-Burke-Ernzerhof\cite{perdew1996generalized} (PBE) generalized gradient approximation (FLO-PBE) and the SCAN\cite{sun2015strongly} functional (FLO-SCAN), in addition to FLO-LDA. They found that using FLO-DFA improved performance for barrier heights in the same order as for the uncorrected functionals, i.e., that SCAN (FLO-SCAN) performed better than PBE (FLO-PBE), and that the latter performed better than LDA (FLO-LDA).  One focus of that work was to understand the success of density-corrected DFT\cite{sim2018quantifying,kim2013understanding,wasserman2017importance} based on self-consistent Hartree-Fock (HF) densities (HF-DFT) in improving barrier heights.  The idea underlying HF-DFT is that errors made by a DFA can arise from problems with the self-consistent DFA density and can be mitigated by evaluating the DFA on the exact density, or a close proxy, typically the self-consistent Hartree-Fock (HF) density.  Kaplan et al. concluded that HF-DFT improves results for BH76 due to a cancellation of errors.\cite{Kaplan2023}  The proxy HF density over-localizes the density in a manner that pushes up the transition state (TS) energy relative to the reactants (R) or products (P), resulting in improved barrier heights. A related finding\cite{Kaplan2023} is that the self-consistent FLO-SCAN density is closer than HF to the exact density for these reactions.  A still more recent study of the BH76 reactions (and other cases where electron self-interaction is expected to be important) by Hernandez et al.\cite{hernandez2023new} also found that HF-DFT mainly works via error cancellation. %and the HF density is not a good proxy for the exact density. 
%In Ref.\citenum{Kaplan2023} the authors determined that the FLO-SCAN density is closer to an accurate reference density than the HF density for the TS of these reactions.  An important implication is that if the FLOSIC SCAN density can be viewed as close to exact for these reactions, then all improvements to the reaction barriers in FLOSIC SCAN calculations stem from the SIC part of PZSIC, in other words, that the corrections are functional-driven.  

Focusing on the question of how self-interaction corrections (SIC) improve reaction barriers, Shukla and co-workers\cite{shukla2023} analyzed the FLO-LDA and LSIC results for the BH76 barriers\cite{mishra2022study} on an orbital by orbital basis, distinguishing between the participant orbitals, orbitals associated with bonds that are being formed or broken in a reaction,  and the remaining spectator orbitals. The analysis identified participant orbitals, particularly stretched bond orbitals that are shared over multiple atoms in the TS, as the largest contributors to the barrier corrections.   

Several questions motivate the current work.  One involves comparing and contrasting SIC to reaction barriers for functionals on different rungs of Jacob's ladder\cite{schmidt2001jacobs} of DFAs. %in FLO-PBE and FLO-SCAN calculations to those in FLO-LDA.  
Kaplan et al.\cite{Kaplan2023} reported that PZSIC-related corrections reduce mean absolute deviations (MAD) from reference values for barrier heights by nearly 13 kcal/mol in FLO-LDA, but by only 5 kcal/mol in FLO-SCAN.  Because of this large difference, one can ask whether the mechanism of the correction is the same in the two methods.  Does the stretched bond contribution also dominate the SIC-based correction for FLO-SCAN?  If so, why are the corrections smaller?  A second question involves how self-interaction corrections related to the stretched bonds evolve over the course of a reaction.  In a recent study\cite{schulze_2023}, the PZSIC-LDA total energy obtained in FLO-LDA calculations was tracked along the reaction pathway for the Diels-Alder reaction.  The PZSIC energy has a relatively large, but narrow peak centered at the TS geometry that is missing from the analogous plot of the DFT energy.  This suggests that  corrections to the reaction barrier are due to changes in the electronic structure that occur only very near the TS.  We test this idea here by monitoring the evolution of the SIC energy associated with participant and spectator orbitals along approximate reaction pathways for four representative reactions.  A final question concerns the effect of using FLOSIC bond lengths for the R, TS, and P species for predicting barrier heights. To date, all studies of the performance of FLOSIC for reaction barriers have used reference coordinates obtained from high-level quantum chemistry calculations.  But previous calculations\cite{johnson_2017,trepte_2019} have shown that bond lengths predicted in FLO-LDA are up to 4\% shorter than LDA values so using FLOSIC-derived coordinates could alter calculated barrier heights.  Here we use FLOSIC atomic forces\cite{trepte_2019} to find optimized geometries for R and P and approximate FLOSIC coordinates for the TS.  We show below that using the FLOSIC-based coordinates improves the predicted barriers, but only slightly.  

%We introduce below a quantity we refer to as the XC/H ratio, the ratio of the magnitude of the self-exchange energy of an orbital to its self-Coulomb energy.  Since these quantities have opposite signs, an XC/H ratio equal to one implies the cancellation of self-exchange and self-Coulomb energies for the orbital.  We show that the XC/H ratios can provide significant insight into SIC-related corrections.

For this work, we studied the three reactions that make up the BH6 benchmark set\cite{lynch2003small} of reaction barriers.  These include $T_{4}$: CH$_4$+OH $\rightarrow$ CH$_3$+H$_2$O, $T_{12}$: H + OH $\rightarrow$ O + H$_2$, and $T_{13}$: H + H$_4$S $\rightarrow$ H$_4$ + HS.  (The labels for the reactions are taken from Ref. \citenum{mishra2022study}. ) In addition, we include the $T_{9}$ reaction: F + H$_2$ $\rightarrow$ HF + H. This reaction is of interest due to the substantial contributions from its spectator orbitals to the SIC, as previously observed in Ref. \citenum{shukla2023}. Moreover, the  overall SIC contribution to the reverse barrier is negative for $T_9$.  This is rare and merits further investigation.
%This reaction is of interest because its spectator orbitals were found to yield relatively large contributions to the SIC corrections for this reaction in Ref. \citenum{shukla2023}  It is also interesting because the overall SIC contribution to the reverse barrier is negative, which is rare and worth understanding in more detail.   
The BH6 reactions were originally selected as a convenient subset of BH76 because the error statistics for BH6 are similar for many DFT methods to those for the full BH76 reaction set.  We can confirm that this is also true for the FLOSIC reaction barriers.  Comparing to data for the full BH76 set given in Ref. \citenum{Kaplan2023}, we find that the MADs found for BH6 using FLO-SCAN, FLO-PBE, and FLO-LDA (see below) are within 1.1 kcal/mol of the corresponding values for the full BH76 set for each method.
          
In the next section, we briefly review  the FLOSIC method and the numerical approaches used in this work.  We then present our results and a discussion. In the final section, we summarize the main findings and offer concluding remarks.

%SIC is a crucial aspect of DFT, particularly in calculating reaction barriers. The self-interaction error arises from the incorrect description of electron-electron interactions within traditional DFT methods, leading to inaccurate energy barriers for chemical reactions. By incorporating self-interaction correction techniques into DFT calculations, more reliable and accurate reaction barrier energies can be obtained. Understanding the role of self-interaction correction in DFT calculations specifically for reaction barriers is vital for advancing our knowledge of chemical reactions and enabling more precise computational modeling of complex systems.

% Line breaks in section headings at all levels can be introduced using
% \textbackslash\textbackslash. A blank input line tells \TeX\ that the
% paragraph has ended. 
\section{THEORY AND METHODOLOGY}
The FLOSIC\cite{pederson2014communication} calculations reported here make use of the PZSIC energy functional\cite{perdew1981self} for which the exchange-correlation part of the energy includes an orbital-by-orbital self-interaction correction:
\begin{equation}
    E^{PZSIC-DFA}_{xc}[n_\uparrow, n_\downarrow] = E^{DFA}_{xc}[n_\uparrow, n_\downarrow] + \Sigma_{i\sigma} E^{SIC}[n_{i\sigma}].
\end{equation}
Here $n_\uparrow$, $n_\downarrow$ are total spin densities, and $n_{i\sigma}$ is the density of a single Fermi-L\"owdin orbital (FLO) and the sum is over all the FLOs.  The SIC energy of a FLO is given by
\begin{equation}
    E^{SIC}[n_{i\sigma}] = -E_H[n_{i\sigma}] - E^{DFA}_{xc}[n_{i\sigma},0]
\end{equation}
where $E_H[n_{i \sigma}]$ is the self-Hartree (or self-Coulomb) energy associated with the FLO density and $E^{DFA}_{xc}[n_{i \sigma}]$ is the self-exchange and correlation energy, evaluated using the given DFA.  Here we used the LDA\cite{perdew1992accurate}, PBE\cite{perdew1996generalized}, and SCAN\cite{sun2015strongly}.

All FLOSIC calculations done for this work used the FLOSIC code \cite{FLOSICcode}, which is based on the earlier NRLMOL software\cite{pederson1990variational,jackson1990accurate,porezag2005}.  Both codes utilize an accurate integration mesh\cite{pederson1990variational} and a default basis set\cite{porezag1999optimized} that is comparable to quadruple zeta quality basis sets for evaluating total energies.\cite{akter2020study}  

Minimizing the PZSIC total energy requires determining both the occupied orbital space and the optimal FLOs spanning that space.  The N FLOs are obtained from the occupied space by a transformation that uses a set of parameters known as Fermi orbital descriptors, or FODs, which are simply N points in 3-D space.  Moving the FODs changes the shape of the FLOs and therefore the magnitude of the SIC part of the total energy.  Gradients of the energy with respect to the FODs can be computed\cite{pederson2015fermi} and used with a gradient optimizer such as LBFGS to determine the minimum-energy arrangements of the FODs.  We utilized a two-step scheme\cite{karanovich_2021} to do this.  First, the occupied orbital space is determined for a fixed choice of FODs via an SCF calculation, using the scheme of Ref. \citenum{yang2017full} and a convergence criterion of 1 $\times$ $10^{-6}$ Ha.  FOD forces are checked at the end of the SCF cycle and if the largest force exceeds the convergence criterion of 2 $\times$ $10^{-4}$ Ha/Bohr, the occupied space is frozen and the FODs are optimized.  Updating the FODs changes the effective Hamiltonians for the FLOs (see Ref.\citenum{yang2017full}), so a new SCF calculation is done, freezing the new FOD positions, to find an updated occupied orbital space. The FOD forces are then checked again and the two-step cycle is repeated until the FOD forces meet the convergence criterion at the end of the SCF calculation. 

Atomic geometry optimizations performed for this work utilized FLOSIC atomic forces computed using the method described in Ref.\citenum{trepte_2019} and LBFGS.  Geometries were considered converged when the largest atomic force dropped below 1 $\times$ $10^{-3}$ Ha/Bohr.  %To determine the coordinates of the TS species within FLOSIC, we used a local search method.  First, at the reference TS coordinates we computed the FLOSIC forces.  Then, using a finite difference approach with step sizes of 0.05 Bohr, we compute the FLOSIC Hessian. We then diagonalized the Hessian to find the normal mode frequencies and eigenvectors.  For all four reactions studied here, this yielded one imaginary frequency, as expected.  The predicted TS geometry was then found by moving the atoms in the eigenvector directions according to the size of the atomic gradients projected onto the eigenvectors and the normal mode frequencies.  

\section{RESULTS AND DISCUSSIONS}
\subsection{Barrier heights and reaction energies}
We began by computing the total energies $E$ of the R, TS, and P configurations for each of the four reactions identified in the Introduction.  Using atomic coordinates taken from Ref. \citenum{Goerigk2017},  we computed the energies with the LDA, PBE, and SCAN functionals, and the corresponding FLO-DFA functionals. Barrier heights for the forward ($f$) and reverse ($r$) reactions are computed as,  $\Delta E_{f} = E(TS) - E(R)$, and $\Delta E_{r} = E(TS) - E(P)$.  This gives a total of four forward and corresponding reverse reaction barriers. Computed barriers for these are shown in Table \ref{tab1}, along with reference values from Ref.\citenum{Goerigk2017}. 

The FLOSIC barrier heights can be decomposed into contributions from the DFA and SIC parts of the PZSIC total energy
\begin{equation}
    \Delta E = \Delta E^{DFA@FLO} + \Delta E^{SIC}
\end{equation}
corresponding to differences in the two terms on the right of Eq. 2.  (For the $f$/$r$ reaction barrier, $\Delta $ is the difference between energies for the TS and R/P configurations.)  $\Delta E^{DFA@FLO}$ for the various reactions is also shown in Table \ref{tab1}.  These are obtained by evaluating the DFA functionals at the self-consistent FLO-DFA densities. $\Delta E^{SIC}$ is also shown in Table \ref{tab1}.  $\Delta E^{SIC}$ can be further analyzed into contributions from individual FLOs. This is done by matching the FLOs in the TS to corresponding FLOs in the R or P and taking the appropriate $E^{SIC}[n_{i\sigma},0]$ differences. Following Ref. \citenum{shukla2023}, we define participant orbitals to be those associated with bonds that form or break in the reaction, while the remaining orbitals are spectators.  The individual orbital contributions to $\Delta E^{SIC}$ are typically larger in magnitude for the participant orbitals than for the spectator orbitals.  The largest positive contribution in nearly all cases and for all functionals is due to the stretched bond orbital $\Delta E^{SIC}_{sb}$.  We show this quantity in Table \ref{tab1} as well as the total SIC contribution $\Delta E^{SIC}$.  

For 6 of the 8 reaction barriers shown in Table \ref{tab1},  $\Delta E^{SIC}$ is positive for all functionals. In these cases, $\Delta E^{SIC}$ for the FLO-LDA calculation is significantly larger than for FLO-PBE, and larger for FLO-PBE than for FLO-SCAN.  For these six barriers, the SIC contribution is an average of 1.8 times larger for FLO-LDA than for FLO-SCAN.  For the two remaining reactions, $T_{4}$(r) and $T_{9}$(r), the total SIC contributions are close to zero or negative.  In these cases, the FLO-LDA and FLO-SCAN values are similar and the FLO-PBE values are somewhat more negative. Comparing $\Delta E^{SIC}_{sb}$ and $\Delta E^{SIC}$ values, for the six reactions with positive $\Delta E^{SIC}$, the stretched bond contribution makes up an average of 79\% of the total for FLO-LDA and FLO-PBE and 70\% for FLO-SCAN.  For the reactions with negative total SIC contributions, the stretched bond contribution is nonetheless positive for $T_{4}$(r) for all functionals and also positive for FLO-LDA for $T_{9}$(r).  These findings are consistent with the results found in Ref. \citenum{shukla2023} for FLO-LDA and indicate that the stretched bond orbital contributes most to SIC-related barrier corrections for all three FLO-DFA methods, regardless of their rung on Jacob's ladder. 

The mean absolute deviations (MAD) of the computed forward and reverse barrier heights from the reference values are shown in Table \ref{tab2} for the BH6 reactions ($T_{4}$, $T_{12}$, and $T_{13}$). The overall average for forward and reverse reactions, $\Delta E^{MAD}_{tot}$, is also shown. This corresponds to the BH6 MAD.  Results are shown for each method.  As found earlier by Kaplan \textit{et al.}\cite{Kaplan2023} for the full BH76 set, the MAD is largest for FLO-LDA (5.10 kcal/mol), smaller for FLO-PBE (4.14 kcal/mol) and smallest for SCAN (3.09 kcal/mol).  The values for the uncorrected functionals are considerably larger, but are in the same order, LDA (18.08) > PBE (9.61) > SCAN (7.88), all in kcal/mol.  Comparing the DFA and DFA@FLO MAD values indicates how much of the improvement in the FLO-DFA calculations can be attributed to changes in the self-consistent density.  For FLO-LDA, -PBE, and -SCAN these contributions are 3.25, 3.29, and 1.28 kcal/mol, respectively.  The  portion of the improvement due to $\Delta E^{SIC}$ is then 9.73, 2.18, and 3.59 kcal/mol, for FLO-LDA, -PBE, and -SCAN respectively.  This analysis shows that SCAN and PBE succeed in reducing the MAD for reaction barriers significantly relative to LDA, but the inclusion of SIC reduces them further.  The overall correction for FLO-LDA is particularly dramatic, reducing the MAD by roughly 13 kcal/mol relative to LDA, and resulting in a performance for  barrier height predictions that is nearly on par with FLO-PBE and FLO-SCAN. 

We also show the MADs for reaction energies, $\Delta E_{rxn} = E(P) - E(R) = \Delta E_{f} - \Delta E_{r}$ in Table \ref{tab2}.  The MADs are again ordered LDA > PBE  > SCAN.  The SCAN value (1.56 kcal/mol) shows impressive accuracy, while the LDA value (11.08 kcal/mol) is poor by comparison.  Interestingly, the DFA and DFA@FLO MAD values are close, differing by less than 1 kcal/mol, for all functionals.  It is also interesting that the FLO-DFA MAD values are significantly larger than the DFA values for the SCAN and PBE functionals.
%If SIE is pragmatically defined in terms of differences between DFA and reference values, the following can be said:  SIE is relatively large for reaction barriers for all three DFAs and is significantly reduced by using FLO-DFA.  Part of the error is associated with density changes brought about using FLOSIC and part is due to $\Delta E^{SIC}$.  SIE is especially large for LDA and the FLOSIC correction is mostly due to $\Delta E^{SIC}$.  For reaction energies, SIE is still large for LDA and still significantly improved by FLOSIC, again attributable mainly to $\Delta E^{SIC}$.  But for PBE, and especially for SCAN, there is little SIE in the reaction energies, and using FLOSIC worsens predictions. It is worth reiterating that 
Since the reaction energy is not independent of the forward and reverse reaction barriers, this means that errors in the reaction energy inevitably imply errors in the reaction barriers.  We return to this point later.

\subsection{Evolution of the self-interaction corrections}
%Motivated by the work of Schulze et al. \cite{schulze_2023} that showed a peak in the SIC energy near the TS of the Diels-Alder reaction, we were interested in tracking the development of the SIC energy of participant and spectator orbitals over the course of the reactions studied here.  
To study the evolution of SIC related to participant FLOs over the course of a reaction, we performed FLOSIC calculations for multiple steps along approximate reaction pathways for the reactions under investigation.  We approximated the forward reaction pathway by linear interpolation from the atomic positions in the R molecules to the corresponding positions in the TS.  We separated the molecules in R by 5\AA~ to insure they do not interact strongly.  The reverse pathway was similarly approximated as a linear interpolation from the separated P molecules to the TS.  Ten steps were made for both the forward and reverse reactions.  At each step, a self-consistent FLO-DFA calculation was performed for each functional, independently optimizing FOD positions in each case.  FODs were fully relaxed in each calculation from starting points obtained from the appropriate linear interpolation of the optimized FODs for R, TS, and P.

\begin{figure}
    \centering
    \includegraphics[scale=0.7]{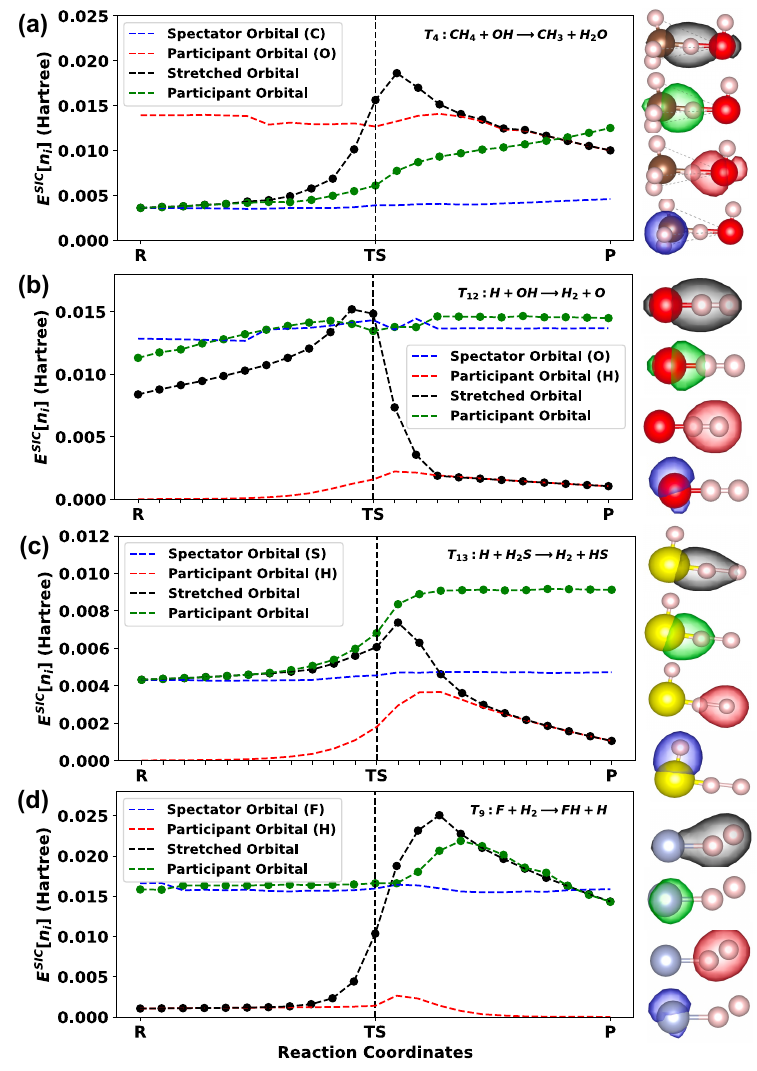}
    \caption{SIC energy, $E^{SIC}[n_i]$, from FLO-SCAN calculations for select spectator, participant, and stretched bond orbitals for four representative reactions. The energies are evaluated at several steps along approximate reaction pathways connecting the reactant (R) species to the transition state (TS), and from the TS to the product (P) species.  Differences in $E^{SIC}[n_i]$ for a given orbital between the TS and the R (P) reflect the contribution of that orbital to the SIC correction of the forward (reverse) reaction barrier. }
    \label{figevo}
\end{figure}

In Figure \ref{figevo} we show $E^{SIC}[n_{i \sigma}]$ for a set of participant and spectator orbitals for each reaction, taken from the FLO-SCAN calculations for steps along the pathways.  At the right of the figure, we show iso-surface plots of the selected orbital densities for the TS configuration.  The energy of what becomes the stretched bond orbital at the TS is shown in black symbols for each reaction.  Data for other participants are shown using green and red symbols.  Blue symbols are used for spectator orbitals.  

It is useful to consider the details of Figure 1 for one of the reactions.  $T_{4}$ can be described as a H transfer from CH$_4$ to OH, producing H$_2$O and CH$_3$.  %The unpaired electron on OH in R has spin up, which implies that the electron transferred with the proton has spin down, allowing the formation of the OH bond, and leaving behind an unpaired spin up electron in CH$_3$.  
In Figure \ref{figevo} (a), the red curve is for the participant oxygen radical (spin up) electron in R that becomes an OH bond orbital in P. The green curve is for a participant FLO that begins as a CH bond orbital in R and ends as an unpaired C radical (spin up) electron in P.  The blue curve corresponds to the SIC energy of a spectator CH bond FLO, i.e. a FLO  associated with the same CH bond throughout the reaction.  As noted above, the black curve is for the stretched bond orbital, which begins as a (spin down) CH bond orbital in R and ends as an OH bond orbital in  P.  At the TS, this  FLO extends over the stretched C-H and O-H bonds.  The curves in the other figures can be interpreted analogously. 

A striking feature of Figure \ref{figevo} is that $E^{SIC}[n_{sb}]$ peaks near the TS in all four reactions.  This is reminiscent of the peak reported by Schulze et al. \cite{schulze_2023} for the PZSIC energy near the TS of the Diels-Alder reaction.  %This is noteworthy because $\Delta E^{SIC}_{sb}$, representing the contribution of the stretched bond orbital to the SIC of a barrier, is determined by the energy difference between its TS and the R or P.
The peak in $E^{SIC}[n_{sb}]$is noteworthy because  $\Delta E^{SIC}_{sb}$, the difference between its energy at the TS and that of the corresponding orbital in the R/P, is the contribution of the stretched bond orbital to the correction of the forward/reverse barrier. The peak in $E^{SIC}_{sb}$ means that this contribution is positive and in 
 Figure \ref{figevo} it can be seen that it is generally large compared to contributions for other orbitals, especially for spectator orbitals.   The plots shown in Figure \ref{figevo} are for the FLO-SCAN calculations, and the analogous plots for FLO-PBE and FLO-LDA (see the Supporting Material (SM) Figures S1, S2, S3, and S4) are qualitatively quite similar.  The stretched bond peak for the $T_{9}$ reaction is shifted to the P side of the TS compared to the other reactions. The reason for this is addressed in the Discussion below.

%The change in $E^{SIC}[n_{sb}]$ values reflects the increase in stretched, multi-bond character near the TS.  Similarly, the large changes in $E^{SIC}$ for the participant orbitals in reaction $T_{4}$ and $T_{13}$ reflect a change in the noded character of the orbitals.  This is shown in Figure S1 in the supplemental material (SM).  The impact of stretched or noded orbitals in SIC calculations is discussed in Ref. \citenum{shahi2019stretched}.   %Semi-local DFAs, such as LDA, PBE, and SCAN, are built on local models of the exchange hole and perform relatively well for compact, nodeless densities, as for the ground state total density of many electron systems. But 

\begin{figure}
   \centering
    \includegraphics[scale=1.0]{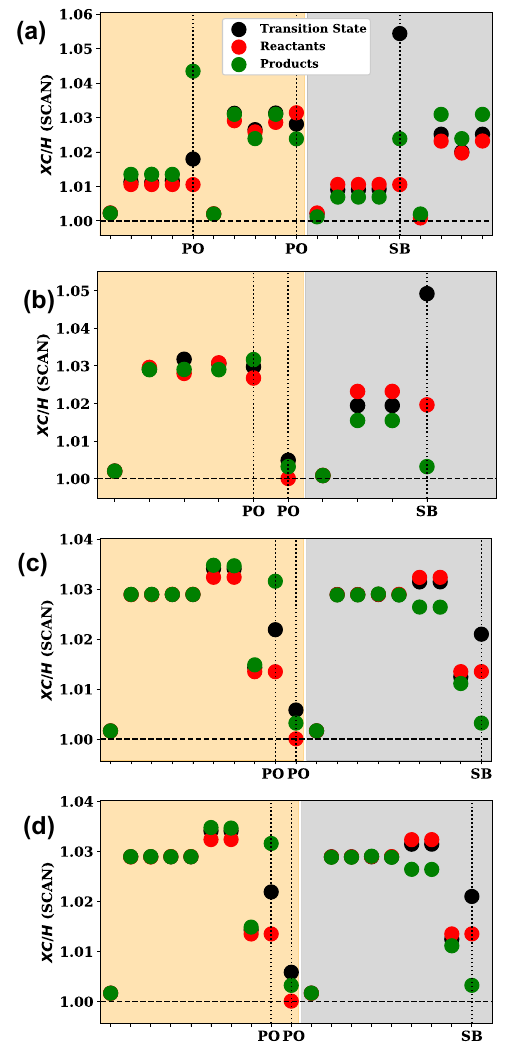}
     
    \caption{The magnitude of the self-exchange-correlation energy (XC) divided by the self-Hartree energy (H) is presented for FLOs in four representative reactions: (a) $T_{4}$: CH$_4$+OH$\rightarrow$CH$_3$+H$_2$O, (b) $T_{12}$: H+OH$\rightarrow$O+H$_2$, (c) $T_{13}$: H+H$_4$S$\rightarrow$H$_4$+HS, and (d) $T_{9}$: F+H$_2$$\rightarrow$FH+H. The XC/H ratios were obtained from self-consistent FLO-SCAN calculations for each reaction. Red, black, and green dots represent XC/H ratios for FLOs in the reactants (R), transition state (TS), and product (P) species, respectively. The horizontal dashed line signifies the reference point where the magnitude of XC equals H. Vertical dotted lines indicate participant (PO) and stretched bond (SB) orbitals involved in the respective reactions. Data for spin-up (down) FLOs are shown on a beige (gray) background. Changes in XC/H for a given FLO reflect changes in the shape of the FLO between the R, TS, and P states. Further discussion is provided in the text.}

%\textcolor{red}{\textbf{PLEASE CHECK THE CAPTION}}
    
    % The magnitude of the self-exchange\textcolor{red}{-correlation} energy (XC) divided by the magnitude of the self-\textcolor{red}{Hartree} energy (H) for FLOs in \textcolor{red}{(a) $T_{4}$: CH$_4$+OH$\rightarrow$CH$_3$+H$_2$O, (b) $T_{12}$: H+OH$\rightarrow$O+H$_2$, (c) $T_{13}$: H+H$_4$S$\rightarrow$H$_4$+HS, and (d) $T_{9}$: F+H$_2$$\rightarrow$FH+H },four representative reactions.  The XC/H ratios were obtained for self-consistent FLO-SCAN calculations for each reaction. XC/H ratios for the corresponding FLOs in the reactants (R), transition state (TS), and product (P) species are shown using red, black, and green dots, respectively. \textcolor{red}{Horizontal dashed line represents the reference line where the magnitude of XC is exactly equal to H. Verticle dotted lines show participant (PO) and stretched bond (SB) orbitals involved in the respective reaction.}   Data for spin-up (down) FLOs are given on the beige (gray) background. \sout{The order of the FLOs is arranged such that values for participant orbitals are at the far right of each background for each reaction}.  Changes in XC/H for a given FLO reflect a change in the shape of the FLO between the R, TS, and P states. See text for further discussion.
    % }
    \label{figxc}
\end{figure}

\begin{table*}[]
    \centering
    \begin{tabular}{ccccccccc} \hline \hline
           Reactions $\rightarrow$ & \multicolumn{2}{c}{$T_4$} & \multicolumn{2}{c}{$T_{12}$} & \multicolumn{2}{c}{$T_{13}$} & \multicolumn{2}{c}{$T_{9}$} \\ 
           \cline{2-3}\cline{4-5} \cline{6-7} \cline{8-9}
           &$\Delta E_f$&$\Delta E_r$&$\Delta E_f$&$\Delta E_r$&$\Delta E_f$&$\Delta E_r$&$\Delta E_f$&$\Delta E_r$ \\ \hline \hline
           LDA           &-23.71&-17.32 &-12.51&-27.28 &-10.64&-17.05&-25.32&-8.58 \\
           LDA@FLO       &-18.96&-12.50 &-9.99 &-22.54 &-9.33 &-15.64&-16.88&-0.79 \\
           LDA@FLO-SCAN  &-18.54&-12.14 &-10.13 &-22.03 &-9.37 &-15.53&-15.82&-0.11 \\
           FLO-LDA       &-1.90 &-12.49 &-1.28 &-4.74  &-2.64 &-7.52 &-4.55 &-7.46 \\
$\Delta E^{SIC}$         &17.06 &0.01   &8.71  &17.80  &6.69  &8.12  &12.33 &-6.67 \\ 
$\Delta E^{SIC}_{SB}$    &11.62 &13.48  &13.12 &10.64  &3.47  &6.06  &8.63  &10.70 \\ \hline
           PBE           &-11.81&-10.55 &-7.30 &-14.91 &-5.13 &-7.94 &-14.57&-9.04 \\ 
           PBE@FLO       &-7.17 &-6.10  &-4.00 &-10.72 &-3.74 &-6.20 &-6.39 &-1.87 \\
           PBE@FLO-SCAN  &-8.28 &-7.02  &-5.21 &-11.95 &-4.42 &-6.84 &-8.11 &-3.48 \\
           FLO-PBE       &6.03  &-10.36 &2.02  &3.14   &1.12  &-2.19 &2.29  &-12.09\\
$\Delta E^{SIC}$         &13.20 &-4.26  &6.02  &13.86  &4.86  & 4.01 &8.68  &-10.22\\ 
$\Delta E^{SIC}_{SB}$    &9.70  &4.94   &5.82  &11.34  &1.50  & 4.32 &7.28  &-3.14 \\  \hline
           SCAN          &-7.91 &-7.75  &-7.74 &-11.09 &-6.63 &-6.13 &-9.39 &-11.61\\
           SCAN@FLO      &-6.05 &-6.21  &-6.37 &-9.27  &-6.22 &-5.50 &-5.66 &-8.52 \\
           FLO-SCAN      &4.56  &-7.01  &-0.72 &1.69   &-2.32 &-2.21 &2.42  &-14.99 \\
$\Delta E^{SIC}$         &10.61 &-0.8   &5.64  &10.96  &3.90  &3.29  &8.08  &-6.47\\ 
$\Delta E^{SIC}_{SB}$    &7.54  &3.52   &4.05  &8.65   &1.09  & 3.14 &5.84  &-2.50 \\ \hline \hline
Reference\citenum{Verma2019}&6.3   &19.5   &10.9  &13.2   &3.9   &17.2  &1.6   &33.8  \\
           \hline \hline
         
    \end{tabular}
    \caption{The calculated errors in the reaction barrier heights, in units of kcal/mol, for both the forward ($\Delta E_f$) and reverse ($\Delta E_r$) directions of four representative reactions from the BH76 dataset, employing various density functionals.  Also shown are $ \Delta E^{SIC}_{SB}$, the contribution of the stretched bond orbital to the SIC correction for each barrier, along with the total SIC correction $\Delta E^{SIC}$. DFA@FLO labels barriers computed from energies evaluated using the uncorrected DFA, but the self-consistent FLO-DFA density. DFA@FLO indicates barriers computed with the DFA functional evaluated on the self-consistent FLO-SCAN density.}
    \label{tab1}
\end{table*}
\begin{table}[]
    \centering
    \begin{tabular}{ccccc} \hline \hline
        Functionals&$\Delta E^{MAD}_f$&$\Delta E^{MAD}_r$ & $\Delta E^{MAD}_{tot}$&$\Delta E^{MAD}_{rxn}$\\ \hline
           LDA         &15.62&20.55& 18.08 & 11.08 \\
           LDA@FLO     &12.76&16.89& 14.83 (13.41) & 10.35 (8.34) \\
           FLO-LDA     &1.94 &8.25 & 5.10 (5.02) & 5.46 (4.81) \\
           PBE         &8.08 &11.14& 9.61 & 4.31 \\
           PBE@FLO     &4.97 &7.67 & 6.32 & 3.69 \\
           FLO-PBE     &3.06 &5.23 & 4.14 & 8.80 \\
           SCAN        &7.43 &8.32 & 7.88 & 1.56 \\ 
           SCAN@FLO    &6.21 &6.99 & 6.60 & 1.66 \\ 
           FLO-SCAN    &2.54 &3.64 & 3.09 & 7.88\\
           \hline \hline  
    \end{tabular}
    \caption{Mean absolute deviation relative to reference values (MAD) for both the forward ($\Delta E^{MAD}_f$) and reverse ($\Delta E^{MAD}_r$) directions of the BH6 subset ($T_4$, $T_{12}$, and $T_{13}$) of the BH76 dataset, employing various density functional approximations DFAs.  The fourth column gives $\Delta E^{MAD}_{tot}=(\Delta E^{MAD}_f+\Delta E^{MAD}_r)/2$, the overall error for the BH6 reactions. The last column shows the MADs in the calculated reaction energies $\Delta E^{MAD}_{rxn}$ for the three reactions. For FLO-LDA, the values in parentheses represent the MADs for calculations that use atomic geometries optimized within FLO-LDA.  Other values correspond to calculations that use reference coordinates.  Rows designated DFA@FLO give values obtained using the uncorrected DFA evaluated on the self-consistent FLO-DFA density.  All values are in kcal/mol.}
    \label{tab2}
\end{table}

\subsection{XC/H ratio}
The exact, but unknown, density functional is self-interaction free.  It is argued in Ref. \citenum{shahi2019stretched} that $E^{SIC}[n_{i \sigma}] = 0$ for the exact functional for any one electron density, or that the self-Hartree energy (H) for $n_{i \sigma}$ is equal to the magnitude of the self-exchange-correlation energy (XC). (The self-correlation energy is zero for the exact functional, as it also is for SCAN).  We define the ratio XC/H as

\begin{equation}
XC/H = \frac{|E^{DFA}_{xc}[n_{i\sigma},0]|}{E_{H}[n_{i\sigma}]}
\end{equation}

XC/H = 1.0 for any orbital using the exact functional, and the closer the XC/H ratio is to 1.0 for an approximate functional, the closer that DFA is to the exact functional in the self-interaction sense.  The XC/H ratio also implies the sign of the SIC energy for a given FLO.  If XC/H < 1, the self-Hartree energy dominates the self-exchange-correlation for that FLO and $E^{SIC}[n_{i \sigma}] < 0$, whereas when XC/H > 1 the opposite is true and $E^{SIC}[n_{i \sigma}] > 0$.

We computed XC/H ratios for all FLOs for the R, TS, and P configurations of all four reactions.  These are shown in Figure \ref{figxc} for the SCAN functional, where  red, black, and green symbols are used for FLOs in the R, TS, and P, respectively. Data for corresponding FLOs are plotted at the same horizontal position and values for the spin up orbitals are plotted on the beige background and for spin down on the grey.  Values for participant orbitals (PO) and for the stretched bond orbital (SO) are indicated by labels on the x-axis and dotted lines.  Analogous plots for LDA and PBE are shown in Figures S1, S2, S3, and S4 in the SM. They are very similar to Figure \ref{figxc}, but the values of XC/H differ.  For SCAN (Figure \ref{figxc}), XC/H ranges from about 1.0 to around 1.06.  For PBE and LDA the corresponding ranges are 0.99 to 1.13, and 0.87 to 0.99, respectively.  This coincides with the fact that $E^{SIC}$ is negative for LDA and positive for SCAN and PBE.  %$E^{SIC}$ for PBE is often positive but can be negative.

The XC/H values are clearly grouped by orbital type in Figure \ref{figevo}.  For example,  XC/H is close to 1.0 in SCAN for 1s core FLOs, (close to 1.00 for PBE and 0.88 for LDA).  This is true for all atom types in the reactions studied here: H, C, O, F, and S.  This shows that XC/H is sensitive to orbital shape, rather than to how compact the orbital is.  %The SIC energies for these 1s FLOs are close to zero in FLO-SCAN.  
CH bond orbitals have XC/H values near 1.01, OH bond orbitals near 1.02, and O lone pair orbitals near 1.03. %The large values are due to the lobed nature of the orbitals.\cite{} (cite the noded orbital paper)
     
Figure \ref{figxc} shows that  XC/H values do not change significantly from the R to TS to P configurations for spectator orbitals.  There is correspondingly little change in $E^{SIC}[n_{i \sigma}]$ for these orbitals and therefore little contribution from one of these orbitals to SIC energy differences for the reaction barriers or the reaction energy.  On the other hand, XC/H values for participant orbitals can change significantly.  These orbitals therefore have a larger impact on SIC contributions to the barriers or the reaction energy.

The change in XC/H is due to changes in both XC and H.  For example, the stretched bond orbital in the TS is less localized than the corresponding orbitals in either R or P.  The magnitude of H is therefore smaller for the stretched bond.  In the $T_{9}$ reaction, H$_{TS}$/H$_{R}$$ = 0.9$, where H$_{TS}$ is the magnitude of the self-Hartree energy in the TS, etc.  For the $T_{4}$ reaction C$_{TS}$/C$_{R}$ = 0.8.  The magnitude of XC also decreases for the less-localized orbitals, but less than for H, resulting in a larger XC/H ratio.  %The difference in XC/H between R and TS or between P and TS, implies a net contribution to $\Delta E^{SIC}$, since a more positive XC/H implies a more positive (or less negative, for LDA) $E^{SIC}$.

  %  \item Evolution of the stretched bonds w.r.t. reaction coordinates. Approximate reaction coordinates by linear interpolation between reactants and transition state (forward) and reactants and transition states (reverse).  Plot Compare SIC energy of %Using Atomic Simulation Environment to interpolate and perform NEB calculation with NRLMOL as a calculator (working on it).
    
   % \item Comparison of the SIC due to stretched bond across LDA, PBE, and SCAN functionals. (Here we may talk about electrons that are strongly localized or delocalized). The contribution of spectator and participating orbitals in SIC-ENERGY across different functionals.
    
  %  \item Since PBE and SCAN functional have some form of inbuilt self-interaction correction. We can benchmark the computational cost vs error-associated barrier using LDA, PBE, or SCAN functionals. 

\subsection{The effect of FLO-SIC-optimized coordinates on reaction energetics}  

All previous studies of barrier heights in FLOSIC\cite{mishra2022study,Kaplan2023,shukla2023} used reference coordinates\cite{Goerigk2017} for the R, P, and TS species.  It is known that bond lengths optimized in FLO-SIC are shorter by a  few percent than reference values\cite{johnson_2017,trepte_2019}.  This raises the question of how FLO-SIC barrier height predictions would be affected if FLO-SIC geometries were used in the calculations.  To address this, we used FLO-SIC atomic forces\cite{trepte_2019} to re-optimize the various atomic geometries for R and P.  In each case, we started from the reference geometries and used the LBFGS gradient optimization scheme to determine the minimum FLO-SIC energy geometries.  The resulting FLOSIC bond lengths are typically about 2\% shorter than the reference values.  A full comparison of reference and FLOSIC bond lengths is given in Table S1 of the SM.

To obtain FLOSIC coordinates for the TS states, we performed a local search for the TS geometries.  This involved first computing the Hessian matrix at the TS reference coordinates (using finite differences of the atomic forces). We then diagonalized the Hessian to determine vibrational frequencies.  In all cases, we found one imaginary frequency.  Next, we used the forces obtained at the reference geometry in conjunction with the Hessian to predict the approximate positions of the atoms at the TS.  We then carried out a new self-consistent FLO-SIC calculation at the predicted TS geometry and re-evaluated the FLO-SIC forces. At the exact TS geometry, all atomic forces vanish, as the TS is a point of unstable equilibrium.  The magnitude of the 3N-dimensional atomic force vector $||F||$ therefore provides a useful check on the approximate TS geometry obtained from the local search.  For $T_{4}$, for example, $||F|| = 0.053$ Ha/Bohr at the reference TS geometry, but only $0.003$ Ha/Bohr for the predicted FLO-LDA geometry.  A comparison of reference and FLOSIC TS geometries are given in Table S2 in SM.

In Table \ref{tab2} we show the MAD for barrier heights and reaction energies using FLO-LDA geometries in parentheses alongside the MADs obtained using the reference geometries. The MAD for barrier heights changes from 5.10 to 5.02 kcal/mol, and for reaction energies from 5.46 to 4.81 kcal/mol when FLOSIC geometries are used in place of reference geometries. Using FLOSIC-optimized coordinates therefore improves agreement with reference values, but only slightly.

\subsection{Discussion}
The results described above are consistent with the analysis of stretched and noded orbitals in SIC given in Ref. \citenum{shahi2019stretched}.  A key conclusion of that work is that semi-local density functionals must overestimate the exchange energy when  evaluated on stretched or noded orbital densities, due to the local nature of the exchange hole on which the DFA's are based.  As discussed in greater detail in that paper, for a stretched or strongly lobed orbital, the fraction of the charge present in each lobe "sees" an exchange hole that integrates to a full electron, whereas the exact hole is delocalized over both lobes.  This implies that the DFA overestimates the size of the self-exchange energy for a lobed or stretched orbital.  This is apparent in the results presented in Figures 1 and 2, particularly for the stretched bond orbitals. The XC/H ratios for these orbitals are larger and $E^{SIC}$ values are more positive than for corresponding orbitals in R and P. Other orbitals with large XC/H values also have a strongly lobed character.  An example is the unpaired electron in CH$_3$ (the green dot marked by the leftmost PO label for the $T_{4}$ majority spin orbitals) that has an XC/H value close to 1.04.  Given the $sp^2$ hybridization of the bonding in this molecule, the radical electron would be expected to be an unhybridized C $p$ state in a DFT calculation. In the FLOSIC calculation, the FLO associated with the unpaired electron hybridizes somewhat with the CH bond orbitals to lower the SIC energy, but the FLO retains a mainly $p$ character and is strongly lobed. (See Figure S5 in the SM)  

As noted above, a compact and node-less 1$s$ orbital has an XC/H ratio very close to 1.0 for the SCAN functional.  It could be said that SCAN is effectively 1-electron self-interaction free for such orbitals.  %(Ironically, for the LDA functional, the XC/H ratio is closest to 1.0 for stretched bond orbitals.)  
On average, FLOs have XC/H ratios much closer to 1.0 in FLO-PBE and FLO-SCAN than in FLO-LDA.  One consequence is that the self-interaction-corrected total energies are much closer to the uncorrected total energies for FLO-SCAN and FLO-PBE. The size of $E^{SIC}$ for FLO-LDA is an order of magnitude larger than for FLO-PBE or FLO-SCAN and somewhat larger for FLO-SCAN than for FLO-PBE.   PBE and SCAN can be said to be more self-interaction free than LDA.  Another consequence is that the size of the self-interaction barrier corrections in these functionals is smaller than in FLO-LDA.

The impact of SIC on barrier heights involves differences in  $E^{SIC}[n_i]$.  %For the barrier heights of interest here, we consider differences for spectator orbitals and participant orbitals.  
SCAN is not self-interaction-free for most orbitals, but if the lobed-ness of an orbital is not appreciably changed in going from R or P to TS, its self-interaction energy will cancel when barrier heights are evaluated.  The XC/H plots in Figure \ref{figxc} show that participant orbitals are more likely to make large contributions to barrier heights, although spectator orbitals can also contribute in a non-negligible way,\cite{shukla2023} in part because there are generally more spectator orbitals than participants.

The peaks in $E^{SIC}$ for the stretched bond orbitals for FLO-SCAN shown in Figure \ref{figevo} are at or very near the TS except for $T_9$, where the peak is displaced toward P.  Given the interpretation of the peak as reflecting the stretched bond character of the orbital, one expects the peak to coincide with the configuration where the orbital is most stretched and  intuitively, this is the TS.  %We cannot say whether the peak should be precisely at the TS.  The TS is a saddle point on the total energy surface, but the energies represented by the curves in Figure \ref{figevo} contribute only to the SIC part of the energy.  Also, the TS defined by the reference coordinates does not coincide with the FLOSIC coordinates of the TS.  
The plots in Figure \ref{figevo} cannot precisely locate the peak relative to the TS.  First, the reaction pathways used to create Figure \ref{figevo} are only approximate.  Also, the pathways are based on reference coordinates instead of FLOSIC coordinates. These issues are likely to affect the apparent position of the peaks relative to the TS in Figure \ref{figevo}.  More importantly, the $T_9$ reaction is unique among the reactions shown in Figure \ref{figevo} in being a very early reaction. The optimal $H_2$ bond length is about 0.74 \AA.  In the TS of $T_9$, the H-H bond distance is only slightly larger, 0.77 \AA.  For comparison, in the $T_{12}$ and $T_{13}$ TS, the H-H bond distance is 0.91 and 1.15 \AA, respectively.  Thus, the geometry of the TS in $T_9$ is much closer to the R than the P.  The position of the stretched bond peak in $T_9$ may simply reflect that the H-H bond is not yet very stretched at the TS configuration.  The displaced position of the peak also explains the negative contribution of the stretched bond orbital to the reverse $T_9$ barrier in FLO-SCAN.  From Figure \ref{figevo}(b) it can be seen that $E^{SIC}[n_{sb}]$ is below the value of the corresponding F-H bond orbital in P.  If the correction were computed at the peak of the $E^{SIC}[n_{sb}]$ curve, the contribution would be positive, as is the case for the other reactions.

Changes in the XC/H values for the orbitals in Figure 2 also portend SIC contributions to reaction energies, $\Delta E_{rxn}$. For LDA, the corrections reduce the MAD in the reaction energies for the four reactions studied here.  But SCAN  predicts these reaction energies very accurately (MAD = 1.56 kcal/mol) and  using FLO-SCAN worsens the predictions (MAD = 7.88 kcal/mol). The problem is clearly not due to changes in the density. The data in Table \ref{tab2} show that the MAD for reaction energies are very similar in the self-consistent DFA calculations and the DFA@FLO calculations for all three functionals.  This is not surprising since the R and P configurations involve molecules at their equilibrium configurations where density-driven errors are not expected.  Instead, the reaction energy error stems from differences in SIC energies for corresponding orbitals in the R and P configurations.  These can arise because of changes in orbital hybridization or the shift of a bond orbital to a radical orbital as foretold by shifts in XC/H values in Figure 2.  Barrier height errors depend on reaction energy errors, since $\Delta E_{rxn} = \Delta E_{f} - \Delta E_{r}$.  Therefore at least some of the small MAD for the BH6 barrier heights that remain for FLO-SCAN (3.09 kcal/mol) must be attributable to the reaction energy errors. %SCAN can be remarkably accurate when describing properties for systems with atoms near equilibrium positions and, FLO-SCAN  

As mentioned in the Introduction, Kaplan et al.\cite{Kaplan2023} investigated the success of density-corrected DFT in computing the barrier heights of the BH76 reactions.  To do this, they introduced three proxies for the exact functional, including the LC-$\omega$PBE range-separated hybrid functional,\cite{vydrov2006assessment} a global half-half hybrid of SCAN and HF exchange, and the FLO-SCAN method.  They judged LC-$\omega$PBE to be the most reliable of these and FLO-SCAN the least.  Despite the differences, all three proxies point to the same conclusion, that density-corrected DFT corrects the BH76 barrier heights due to a cancellation of errors.  Use of the HF density causes a positive density-driven error of the transition state energies that cancels a large, negative functional-driven error of the DFAs.  More recent work by Kanungo et al.\cite{Kanungo2023} reinforces this conclusion.  In that work, the accurate CCSD(T) density for the T9 reaction is inverted to find the corresponding Kohn-Sham orbitals that are then used to evaluate density-driven errors.  They find density driven errors of roughly 1 kcal/mol for both the forward and reverse T9 reactions.

While judged to be the least reliable proxy for the exact functional,\cite{Kaplan2023} FLO-SCAN can conveniently be used to show the small size of the density-driven errors for the barrier heights studied here. Evaluating SCAN on the FLO-SCAN density (the SCAN@FLO-SCAN results in Table \ref{tab2}) scarcely reduces the MAD for SCAN, by only a little more than 1 kcal/mol, from 7.88 to 6.60.  For the full BH76 set, the FLO-SCAN proxy predicts a density driven error for SCAN of -1.97 kcal/mol,\cite{Kaplan2023} which is close to the prediction obtained with the LC-$\omega$PBE proxy, -0.58 kcal/mol. We can also use the FLO-SCAN density to evaluate density errors the LDA (LDA@FLO-SCAN) and PBE (PBE@FLO-SCAN) functionals.  Results for the barrier heights using energies obtained in this way are shown in Table 
\ref{tab1}.  In both cases, these are remarkably close to the DFA@FLO results for these functionals, with the largest differences of around 1 kcal/mol.  For LDA, individual total energies for the R, TS, and P configurations evaluated on the FLO-SCAN density are strikingly close to the LDA@FLO-LDA total energies (see Tables S5-S8 in the supplemental information).  This indicates that LDA does not distinguish differences between the FLO-SCAN and FLO-LDA densities, at least in terms of total energies.  The reverse is less true:  SCAN@FLO-SCAN and SCAN@FLO-LDA total energies differ by up to 4 kcal/mol for R, TS, and P of individual reactions.

\section{Summary}
We have presented a comprehensive  study of how self-interaction corrections using the PZSIC energy functional and the FLOSIC method affect predicted reaction barrier heights and reaction energies. Our analysis encompasses 8 barriers for 4 representative reactions drawn from the BH76 benchmark set and for the LDA, PBE, and SCAN functionals.  As  discussed earlier by Kaplan et al. \cite{Kaplan2023}, we find that the mean absolute deviation of calculated barrier heights relative to accurate reference values decreases in the order LDA > PBE > SCAN for the uncorrected functionals and in the same order for FLO-LDA > FLO-PBE > FLO-SCAN.  We also show that using FLOSIC-optimized atomic coordinates, instead of reference coordinates, for the reactant (R), transition state (TS) and product (P) configurations reduces deviations of FLO-DFA barriers from reference values, but the effect is very small.

A significant portion of our investigation involves orbital-by-orbital analysis of SIC contributions to the barrier heights.  We demonstrate that participant orbitals (orbitals associated with bonds that are breaking or forming in a reaction) play a crucial role in influencing  SIC barrier corrections. Among these, stretched bond orbitals that extend over multiple atoms contribute the most.  This is consistent with the results of Ref. \citenum{shukla2023} for FLO-LDA and extends them to FLO-PBE and FLO-SCAN. Our analysis further reveals that the corrections associated with the stretched-bond orbitals emerge only for configurations along the reaction path near the TS.

We introduced the parameter XC/H, the magnitude of the ratio of the self-exchange-correlation energy of an orbital to its self-Hartree energy. We show that XC/H values cluster by orbital type and are related to the shape of the orbital.  We also argue that an XC/H near 1 indicates relative one-electron self-interaction freedom of a functional for that orbital. By this measure, SCAN is relatively self-interaction-free for orbitals that are not stretched or significantly lobed. Our XC/H analysis is consistent with the understanding of stretched and lobed orbitals presented in Ref. \citenum{shahi2019stretched}.

%We find that FLO-SCAN, FLO-PBE, and FLO-LDA all make sizable errors in reaction energies for the reactions studied here.  In the case of FLO-SCAN and FLO-PBE, the MAD is much larger than for SCAN and PBE.  We point out that errors in reaction energies limit how accurate the forward and reverse barriers can be.  Finally, we also showed that using FLOSIC-optimized coordinates for computing reduces deviations from reference values slightly, but the effect is very small.

The SCAN functional is very accurate for many properties when atoms are at or near their equilibrium positions.  For example, SCAN gives reaction energies, based on the energies of the equilibrium R and P configurations, within 1.5 kcal/mol of reference values for the reactions studied here.  But the results of this work reinforce the earlier conclusion\cite{shahi2019stretched} that SCAN must fail for stretched/lobed densities because of its semi-local nature. 
As shown here and elsewhere,\cite{mishra2022study,Kaplan2023} PZSIC can largely correct self-interaction-related errors associated with stretched bonds in a reaction TS.  Yet our results also give insight into the limitations of PZSIC when applied in contexts where the effects of electron self-interaction are not expected to be large.  Differences in $E^{SIC}[n_{i \sigma}]$ for corresponding orbitals in R and P configurations are heralded by related changes in XC/H values and result in a significant worsening of predicted reaction energies.  Such effects appear unavoidable, given changes in hybridization, etc. of the orbitals over the course of a reaction.  Since forward and reverse reaction barrier heights are not independent of reaction energies, the worsening of reaction energies in FLO-SCAN must limit its performance in predicting barriers.  The MAD shown in Table \ref{tab2} for FLO-SCAN applied to the BH6 reaction set may be the smallest possible for PZSIC when used with a semi-local functional.  It is known from the work of Santra and Perdew\cite{santra2019perdew} that PZSIC-LDA is not exact in the uniform density limit.  This led to the idea of locally scaling the self-interaction correction, to mute the corrections for slowly-varying densities.\cite{zope2019step}  The resulting LSIC-LDA method performs better than PZSIC-LDA in many settings,\cite{zope2019step,yamamoto2020improvements,akter2020study,romero2021local} but is less successful when used with PBE or SCAN due to gauge inconsistencies.\cite{bhattarai2020step}  
% This limits the accuracy of predicted reaction barriers.  It seems a promising strategy to limit self-interaction corrections for SCAN to situations where SIE is prominent, such as in the calculation of reaction barriers.  Our results suggest that a possible approach may be distinguishing between noded and stretched orbitals when applying SIC.  
We hope that insights drawn from the present work help lead to improved SIC formulations that extend the accuracy of SCAN to situations where self-interaction errors are large.

\section{Associated Content}
{\bf Supporting Information}
The Supporting Information is available free of charge online.

\section{Author Information}

\begin{acknowledgments}
It is a pleasure to participate in this festschrift honoring Professor John Perdew on the occasion of his 80th birthday. John has had a remarkable career at the forefront of the field of density functional theory.  His contributions to the development of accurate and widely used density functional approximations are second to none.  The authors are happy to acknowledge several helpful conversations with John regarding the work described in this paper and about density functional theory in general.  We wish him continued success in his quest to develop a density functional approximation of everything.

We would also like to thank Professor Karl Johnson and Priyanka Shukla for their support at the beginning of this project.

This work was supported by the U.S. Department of Energy, Office of Science, Office of Basic Energy Sciences, as a part of the Computational Chemical Sciences Program under Award No. DE-SC0018331.  Calculations were carried out at the high-performance computing center (HPCC) of the Institute for Cyber-Enabled Research (ICER) at Michigan State University using resources owned by Central Michigan University.
\end{acknowledgments}

\section*{Data Availability Statement}

The data that supports the findings of this study are available within the article and its supplementary material.

% \nocite{*}
\bibliography{references}

\end{document}